\documentclass[twocolumn,english,aps,prc,showpacs]{revtex4}
\usepackage[T1]{fontenc}
\usepackage[latin9]{inputenc}
\usepackage{longtable}
\usepackage{amsmath}
\usepackage{graphicx}
\usepackage{amssymb}

\makeatletter

\providecommand{\tabularnewline}{\\}

\usepackage{longtable}

\usepackage{booktabs}

\usepackage{esint}



\makeatother

\usepackage{babel}

\begin{document}

\preprint{This line only printed with preprint option}

\title{On the $^{14}$C($\alpha$,$\gamma$) reaction rate.}

\author{E.D. Johnson}

\email{edj04@fsu.edu}

\affiliation{Department of Physics, Florida State University, 32306 FL}

\author{G.V. Rogachev}

\email{grogache@fsu.edu}

\affiliation{Department of Physics, Florida State University, 32306 FL}

\author{J. Mitchell}

\affiliation{Department of Physics, Florida State University, 32306 FL}

\author{L. Miller}

\affiliation{Department of Physics, Florida State University, 32306 FL}

\author{K.W. Kemper}

\affiliation{Department of Physics, Florida State University, 32306 FL}

\begin{abstract}
The $^{14}$C($\alpha$,$\gamma$) reaction rate at temperatures below
0.3 GK depends on the properties of two near threshold resonances
in $^{18}$O, the 1$^{-}$ at 6.198 MeV and the 3$^{-}$ at 6.404
MeV. The $\alpha$+$^{14}$C Asymptotic Normalization Coefficients
(ANCs) for these resonances were determined using the $\alpha$-transfer
reactions $^{14}$C($^{7}$Li,t) and $^{14}$C($^{6}$Li,d) at sub-Coulomb
energies. The $^{14}$C($\alpha$,$\gamma$) reaction rate at low
temperatures has been evaluated. Implications of the new reaction
rate on the evolution of accreting helium white dwarfs and on the
nucleosynthesis of low mass stars during the asymptotic giant branch
(AGB) phase are discussed. 
\end{abstract}


\maketitle

\section{introduction\label{sec:introduction}}

The $^{14}$C($\alpha$,$\gamma$)$^{18}$O reaction has been suggested
to play an important role in several astrophysical environments. It
was pointed out by Mitalas \cite{Mitalas1974} that electron capture
on $^{14}$N can produce $^{14}$C in the degenerate helium core of
low mass stars. The $\alpha$-capture on $^{14}$C then follows. It
was suggested in \cite{Kaminisi1975} that the $^{14}$N(e$^{-}$,$\nu$)$^{14}$C($\alpha$,$\gamma$)$^{18}$O
reaction (NCO reaction) may trigger the helium flash in the core of
low mass stars earlier than the $3\alpha$ reaction. The significance
of the NCO reaction for the evolution of low mass stars was also considered
by Spulak \cite{Spulak1980} who concluded that it is not effective
compared to the 3$\alpha$ reaction. A contradicting result was obtained
by Kaminisi and Arai \cite{Kaminisi1983} who used the updated (but
still very uncertain) $^{14}$C($\alpha$,$\gamma$) reaction rate
and concluded that the NCO reaction is dominant for igniting the helium
flash. While the influence of the NCO reaction on the evolution of
low mass stars in the red giant phase is still uncertain it was firmly
established by Nomoto and Sugimoto \cite{Nomoto1977}, and later by
Hashimoto, et al. \cite{Hashimoto1986}, that the NCO reaction may
trigger the helium flash in accreting helium white dwarfs at a lower
temperature and density than the $3\alpha$ reaction. However, this
conclusion is rather sensitive to the actual value of the $^{14}$C($\alpha$,$\gamma$)
reaction rate at temperatures between 0.03 and 0.1 GK.

The $^{14}$C($\alpha$,$\gamma$) reaction is also important for
production of $^{19}$F in asymptotic giant branch stars. Observations
by Jorissen, et al., \cite{Jorissen1992} show enhanced fluorine abundance
in the atmosphere of K, M, MS, S, SC, and C asymptotic giant branch
(AGB) stars. This finding indicates that $^{19}$F is produced in
these stars and its abundance can be used to constrain the properties
of the AGB models. The path for $^{19}$F production in an AGB star,
as proposed by Jorissen \cite{Jorissen1992}, is rather complex, occurring
in the He intershell of the AGB star. Neutrons from the s-process
neutron generator reaction, $^{13}$C($\alpha$,n), can be captured
by $^{14}$N, enhanced from the previous CNO cycle. This leads to
production of $^{14}$C and protons through the $^{14}$N(n,p)$^{14}$C
reaction. Then $^{18}$O is produced by the $^{14}$C($\alpha$,$\gamma$)
reaction, or, alternatively by $^{14}$N($\alpha$,$\gamma$)$^{18}$F($\beta$)$^{18}$O.
The presence of protons and $^{18}$O isotopes in the He intershell
simultaneously allows for the $^{18}$O(p,$\alpha$)$^{15}$N reaction
to occur and subsequently $^{15}$N($\alpha$,$\gamma$)$^{19}$F
capture. Competing reactions that reduce the final abundance of $^{19}$F
are $^{19}$F($\alpha$,p), $^{18}$O($\alpha$,$\gamma$)$^{22}$Ne
and $^{15}$N(p,$\alpha$)$ $$^{12}$C. A detailed investigation
of the $^{19}$F production in AGB stars has been performed recently
by Lugaro, et al., \cite{Lugaro2004} and it was determined that the
major uncertainties in the production of $^{19}$F are associated
with the uncertainties in the $^{14}$C($\alpha$,$\gamma$)$^{18}$O
and $^{19}$F($\alpha$,p)$^{22}$Ne reaction rates. Addressing uncertainties
of the $^{14}$C($\alpha$,$\gamma$) reaction rate at temperatures
relevant for helium flashes in the accreting helium white dwarfs and
nucleosynthesis in the asymptotic giant branch stars is the main goal
of this work.

\begin{figure}
\includegraphics[width=0.9\columnwidth]{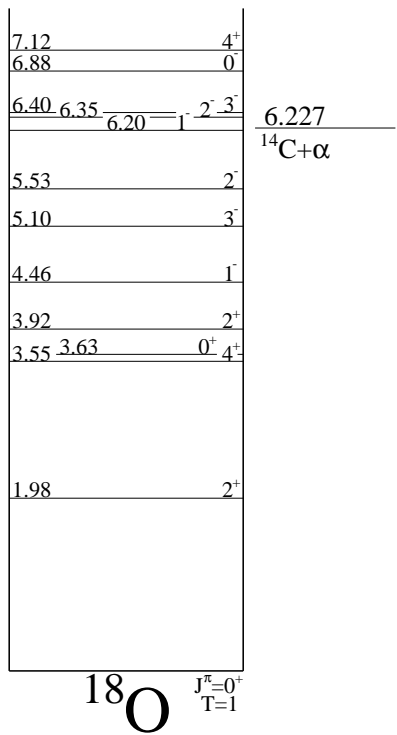}

\caption{\label{fig:18Olevels}Partial level scheme of $^{18}$O.}

\end{figure}

Direct measurements of the $^{14}$C($\alpha$,$\gamma$) reaction
cross section are only available for energies above 880 keV in c.m.
\cite{Gorres1992}. The relevant temperature range for accreting
helium white dwarfs is between 0.03 and 0.1 GK and for $^{19}$F nucleosynthesis
in AGB stars it is $\sim$0.1 GK. The Gamow window for the $^{14}$C($\alpha$,$\gamma$)
reaction at these temperatures corresponds to an energy range between
50 and 250 keV. At these energies the cross section for the $\alpha$
capture reaction on $^{14}$C is too low to be measured directly and
has to be extrapolated from the higher energy data. However, near
threshold resonances in $^{18}$O may have a very strong influence
on this extrapolation. $^{}$The level scheme of $^{18}$O is shown
in Fig. \ref{fig:18Olevels}. Three states with excitation energies
close to the $\alpha$ decay threshold at 6.227 MeV are known in $^{18}$O.
These are the 1$^{-}$ at 6.198 MeV, the 2$^{-}$ at 6.351 MeV and
the 3$^{-}$ at 6.404 MeV. The 2$^{-}$ state at 6.351 MeV is an unnatural
parity state and cannot contribute to the $^{14}$C($\alpha$,$\gamma$)
reaction at any significant level. The 1$^{-}$ subthreshold state
at 6.198 MeV can only contribute to the $^{14}$C($\alpha$,$\gamma$)
reaction through its high energy tail, therefore its contribution
is expected to be relevant only at the lowest energy. Finally, the
3$^{-}$ state at 6.404 MeV is 177 keV above the $\alpha$-decay threshold
- right in the middle of the energy range of interest.

The strength of a resonance is defined by\begin{equation}
\omega\gamma=\frac{2J_{R}+1}{(2j_{t}+1)(2j_{p}+1)}\frac{\Gamma_{\alpha}\Gamma_{\gamma}}{\Gamma},\end{equation}

\noindent where $\Gamma$, $\Gamma_{\alpha}$ and $\Gamma_{\gamma}$
are the total, partial $\alpha$ and partial $\gamma$ widths and
$J_{R}$, $j_{t}$ and $j_{p}$ are spins of the resonance, the target
and the projectile. At 177 keV the partial $\alpha$-width is much
smaller than the partial $\gamma$ width ($\Gamma_{\alpha}\ll\Gamma_{\gamma}$),
therefore the resonant $\alpha$-capture cross section is determined
entirely by the partial $\alpha$-width with resonance strength calculated
as $\omega\gamma\approx7\times\Gamma_{\alpha}$ for the 3$^{-}$ state
at 6.404 MeV in $^{18}$O. The partial $\alpha$-width of the 3$^{-}$
state is not known. This leads to several orders of magnitude uncertainty
of the $^{14}$C($\alpha$,$\gamma$) reaction rate at $\sim0.1$
GK. The subthreshold 1$^{-}$ state at 6.198 MeV also contributes
to the reaction rate uncertainty at the lowest temperatures.

Information on the partial $\alpha$ widths of resonances in $^{18}$O
can be extracted using the direct $\alpha$-transfer reactions $^{14}$C($^{6}$Li,d)
and $^{14}$C($^{7}$Li,t). Usually these reactions are performed
at energies of 5-15 MeV/A and $\alpha$ spectroscopic factors (S$_{\alpha}$)
for the states of interest are determined from the Distorted Wave
Born Approximation (DWBA) analysis of the reaction cross sections.
The partial $\alpha$ width is then calculated as $\Gamma_{\alpha}=S_{\alpha}\times\Gamma_{SP}$,
where $\Gamma_{SP}$ is the $\alpha$ single particle width.

\begin{eqnarray}
\Gamma_{SP} & = & 2P_{\ell}(kR)\gamma_{sp}^{2}\label{eq:width}\\
P_{\ell}(kR) & = & \frac{kR}{F_{\ell}^{2}(k,R)+G_{\ell}^{2}(k,R)}\label{eq:pen}\\
\gamma_{sp}^{2} & = & \hbar^{2}/\mu R^{2},\label{eq:redwidth}\end{eqnarray}

\noindent where $P_{\ell}(kR)$ is a penetrability factor determined
by the Coulomb regular and irregular functions $F_{\ell}(k,R)$ and
$G_{\ell}(k,R)$, $\mu$ is a reduced mass and $k$ is a wave number.
R is a channel radius - $R=r_{0}(\sqrt[3]{A_{1}}+\sqrt[3]{A_{2}})$,
where $r_{0}=1.2-1.4$ fm. The reduced width, $\gamma_{sp}$, in equation
\eqref{eq:redwidth} corresponds to the reduced width in a square-well
potential. The single particle width can be calculated more accurately
using a realistic Woods-Saxon potential. As is well known, the spectroscopic
factor extracted from an $\alpha$-transfer reaction depends on the
parameters of the optical potentials used to describe the wave functions
of relative motion in the entrance and exit channels in the DWBA approach
and on the shape of the form factor potentials and the number of nodes
in the model core-cluster wave function. However, for astrophysical
calculations the asymptotic normalization coefficient (ANC) is needed
instead of the spectroscopic factor, and if the transfer reaction
is performed at sub-Coulomb energies, then the parametric dependence
of the DWBA calculations is greatly reduced. This approach has been
applied before in Refs. \citet{Brune1999,Johnson2006} where the $^{12}$C($\alpha$,$\gamma$)
and $^{13}$C($\alpha$,n) reaction rates due to near threshold resonances
in $^{16}$O and $^{17}$O were evaluated. In this work we measured
the cross sections of the $^{14}$C($^{7}$Li,t) and $^{14}$C($^{6}$Li,d)
$\alpha$-transfer reactions at sub-Coulomb energies and determined
the ANCs of the 1$^{-}$ and 3$^{-}$ states in $^{18}$O. The discussion
is structured as follows: the experimental technique is discussed
in the section \ref{sec:Experiment}, the extraction of the ANCs and
associated uncertainties are described in section \ref{sec:Analysis},
the new $^{14}$C($\alpha$,$\gamma$) reaction rate and it's astrophysical
implications are discussed in section \ref{sec:rate}, and the conclusions
are in section \ref{sec:Conclusions}.

\section{Experiment\label{sec:Experiment}}

The sub-Coulomb $\alpha$-transfer reactions $^{6}$Li($^{14}$C,d)
and $^{7}$Li($^{14}$C,t) were performed at the John D. Fox Superconducting
Accelerator Laboratory at Florida State University. The use of inverse
kinematics, $^{14}$C beam and $^{6,7}$Li target, has several advantages.
First, it eliminates the background associated with ($^{6,7}$Li,d/t)
reactions on the unavoidable $^{12}$C admixture to the $^{14}$C
target. Second, inverse kinematics provides a boost to the light recoils,
which is essential for the detection of the reaction products, since
the reaction is performed at low sub-Coulomb energies.

\begin{figure}
\includegraphics[width=1\columnwidth]{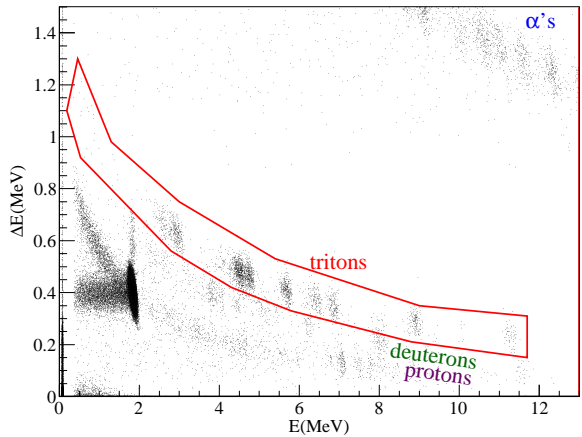}

\caption{\label{fig:Triton dE-E} (Color online) Particle identification spectra
in $\Delta$E-E silicon telescopes from $^{7}$Li($^{14}$C,t) run.
The x and y axes are energies in MeV deposited by the charged particles
in the E and $\Delta$E detectors, respectively. The intense group
of protons at 2 MeV (2.4 MeV total energy) corresponds to elastic
scattering of $^{14}$C by the hydrogen contamination in the $^{7}$Li
target. Tritons from the $^{7}$Li($^{14}$C,t) reaction are highlighted. }

\end{figure}

Five silicon $\Delta$E-E telescopes were used to detect the light
recoils (deuterons/tritons) produced by the $\alpha$-transfer reaction.
The $\Delta$E detectors ranged in thickness from 15 to 25 $\mu m$,
while the E detectors were 500-1000 $\mu m$ thick. Particle identification
was performed using the standard $\Delta$E-E technique. A sample
of the 2D particle identification plot is shown in Figure \ref{fig:Triton dE-E},
where it can be seen that the separation of the different isotopes
of hydrogen is sufficient for reliable identification.

The lithium targets were prepared and transferred to the scattering
chamber in a sealed container under vacuum in order to prevent oxidation.
The thickness of the lithium targets was $\approx20$ $\mu g/cm^{2}$
and they were prepared on a Formvar backing. An accurate determination
of the target thicknesses and detector solid angles was performed
using elastic scattering of protons. The p+$^{6,7}$Li elastic scattering
cross section at 95$^{\circ}$ with proton beam energy of 6.868 MeV
was measured previously with a 3$\%$ accuracy in \citet{Bing70},
and this result allowed the product of the target thickness and the
telescope solid angles to be determined by placing the telescopes
at 95$^{\circ}$ with respect to the beam axis one by one and measuring
p+$^{6,7}$Li elastic scattering with each $ $lithium target.

\begin{figure}
\includegraphics[width=1\columnwidth]{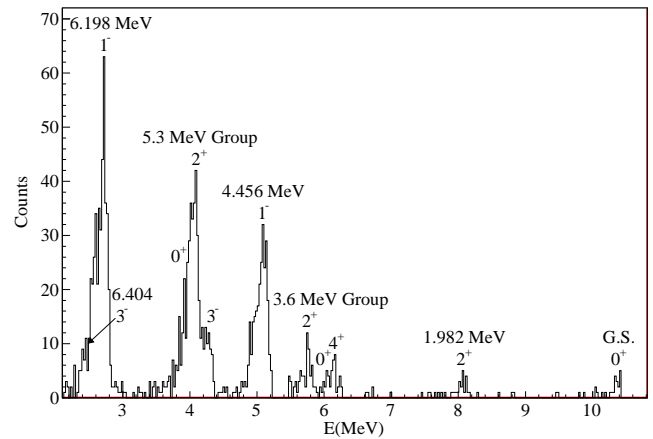}

\caption{\label{fig:Spectrum-of-deuterons}Spectrum of deuterons from the $^{6}$Li($^{14}$C,d)
reaction, measured by the 8$^{\circ}$ telescope at 8.8 MeV $^{14}$C
beam energy. The energy of the deuterons in the laboratory reference
frame is shown on the x-axis. The $^{18}$O states populated in this
reaction are labeled. }

\end{figure}

For energies below the Coulomb barrier the angular distribution for
a direct transfer reaction has a peak at back angles, $180^{\circ}$
in the c.m. frame which corresponds to a peak at $0^{\circ}$ in the
lab frame in inverse kinematics. Therefore, to measure the reaction
cross section in the region of its maximum, the detectors were placed
near $0^{\circ}$ in the lab frame. The detector nearest to $0^{\circ}$
was shielded from Rutherford scattering of the carbon beam on the
lithium target with a 5 $\mu m$ Havar foil in front of the telescope.
This foil was thick enough to stop all of the carbon beam, but thin
enough so that energy losses of the deuterons and tritons were small.

The $^{14}$C beam energy is determined by the condition that the
reaction c.m. energy should be below the Coulomb barrier in both the
entrance and exit channels. The $^{14}$C beam energies of 8.8 MeV
for the $^{6}$Li target (2.64 MeV in c.m.) and 11.5 MeV for the $^{7}$Li
target (3.83 MeV in c.m.) meet this requirement for population of
near $\alpha$ threshold states in $^{18}$O.

\section{Analysis of the $\alpha$-transfer cross sections and ANCs\label{sec:Analysis}}

\begin{figure}
\includegraphics[width=1\columnwidth]{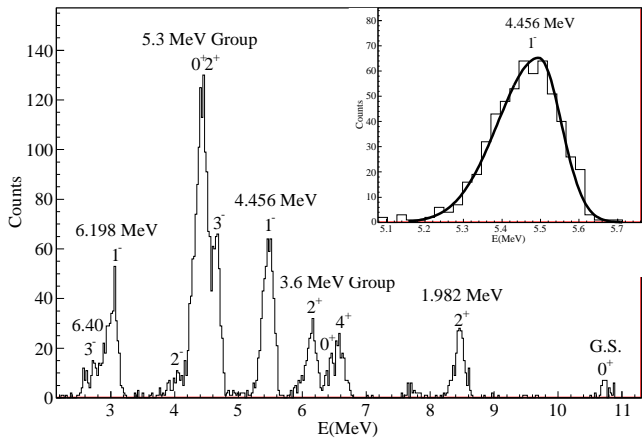}

\caption{\label{fig:Spectrum-of-tritons}Spectrum of tritons from the $^{7}$Li($^{14}$C,t)
reaction, measured by the 8$^{\circ}$ telescope at 11.5 MeV $^{14}$C
beam energy. The energy of the tritons in the laboratory reference
frame is shown on the x-axis. The $^{18}$O states populated in this
reaction are labeled. The inset shows the double Gaussian fit of the
well resolved 1$^{-}$ state at 4.456 MeV which was used to determine
the energy resolution of the experimental setup.}

\end{figure}

Spectra of deuterons from the $^{6}$Li($^{14}$C,d) reaction and
tritons from the $^{7}$Li($^{14}$C,t) reaction at 8.8 MeV and 11.5
MeV of $^{14}$C beam energy, respectively, are shown in Figs. \ref{fig:Spectrum-of-deuterons}
and \ref{fig:Spectrum-of-tritons}. Peaks in the spectra correspond
to the states or groups of states in $^{18}$O and all of them can
be identified with the known states in $^{18}$O. The near threshold
states of astrophysical interest for this work are the 1$^{-}$ at
6.198 MeV and the 3$^{-}$ at 6.404 MeV. Peaks which correspond to
these states appear at 2.5 and 3.0 MeV in Figs. \ref{fig:Spectrum-of-deuterons}
and \ref{fig:Spectrum-of-tritons}, respectively. The experimental
resolution of our setup in the c.m. was 120 keV (FWHM), therefore
these two states partially overlap. The following steps were performed
in order to reduce the ambiguity in the extraction of cross sections
from the measured spectra. The experimental resolution was accurately
determined from the width of the peak which corresponds to the 1$^{-}$
state at 4.456 MeV. This state is separated from the other $^{18}$O
states by more than 500 keV and is therefore well resolved in our
spectra. It was observed that the shape of the 1$^{-}$ peak is slightly
asymmetric with a characteristic {}``tail'' at the lower energy
end (see the inset in Figure \ref{fig:Spectrum-of-tritons}) probably
due to radiation damage in the detectors. This detector response was
then used in the further analysis by modeling it with a Gaussian function
which has larger sigma below the centroid energy and lower sigma above
the centroid energy (110 and 60 keV in the Lab. frame respectively)
. All the peaks in the spectra are then fitted using this lineshape.

\begin{figure}
\includegraphics[width=1\columnwidth]{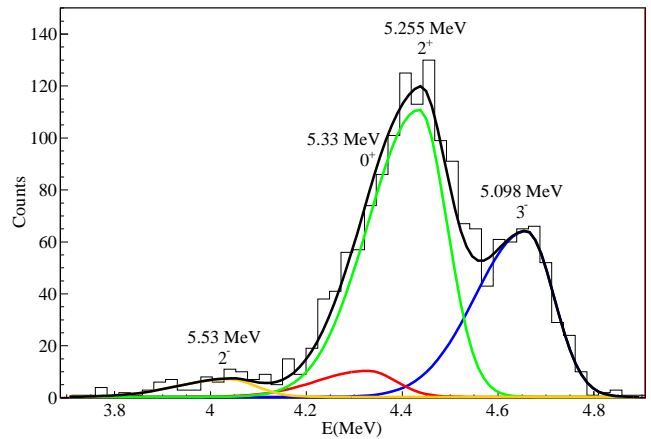}

\caption{\label{fig:2-state}(Color online) Four state fit of the $4.5$ MeV
group of tritons from the $^{7}$Li($^{14}$C,t) reaction. This fit
was performed to determine the strength of the unnatural parity 2$^{-}$
state at 5.53 MeV which was used to evaluate the contribution of the
compound nucleus mechanism to the measured cross sections.}

\end{figure}

Three states can contribute to the broad peak observed at $\sim3$
MeV in the triton and $\sim2.5$ MeV in the deuteron spectra. These
are the aforementioned 1$^{-}$ and 3$^{-}$ at 6.198 MeV and 6.404
MeV but also the 2$^{-}$ at 6.35 MeV. While this last state is an
unnatural parity state and cannot be populated in a direct $\alpha$-transfer
reaction it can still contribute through the compound nucleus reaction
mechanism. The magnitude of this contribution can be estimated using
the 2$^{-}$ state at 5.53 MeV as a benchmark. This state is clearly
visible in the group of states at $\sim4.5$ MeV in the spectrum of
tritons from the $^{7}$Li($^{14}$C,t) reaction (see Figure \ref{fig:2-state}).
Four states contribute to this peak. These are the 2$^{-}$ at 5.53
MeV, the 0$^{+}$ at 5.336 MeV, the 2$^{+}$ at 5.255 MeV and the
$3^{-}$ at 5.098 MeV. The four state fit of the peak is shown in
Figure \ref{fig:2-state}. There is also an unnatural parity 3$^{+}$
state at 5.378 MeV, but its contribution is expected to be very small
and since it has an excitation energy close to the 0$^{+}$ its possible
contribution to the peak can be taken into account by overestimating
the strength of the 0$^{+}$ state. Note that there are only four
free parameters in this fit - the strengths of the four states. The
positions of the states were fixed from the known values and the width
is determined by the experimental shape as modeled above. In this
way, the cross section for population of the 2$^{-}$ state at 5.53
MeV was determined. Assuming that only the compound nucleus mechanism
is responsible for population of the unnatural parity 2$^{-}$ state
at 5.53 MeV one can determine the cross section for population of
the 2$^{-}$ state at 6.35 MeV using the Hauser-Feshbach formalism
and scaling it to the 2$^{-}$ state at 5.53 MeV.

\begin{figure}
\includegraphics[width=1\columnwidth]{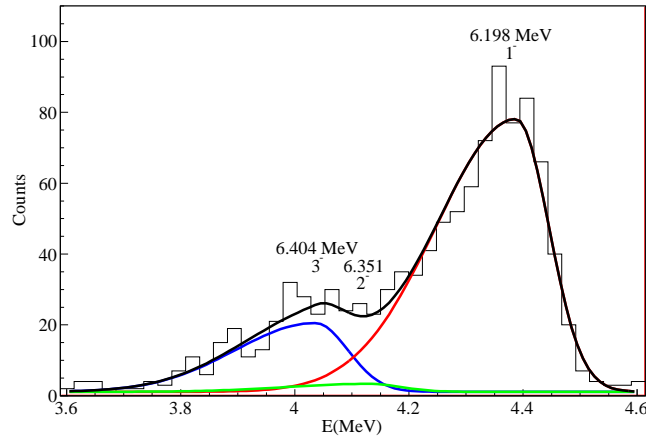}

\caption{\label{fig:main-fit}(Color online) Fit of the triton spectrum from
the $^{7}$Li($^{14}$C,t) reaction in the vicinity of 6.3 MeV of
$^{18}$O excitation energy. The black curve is the sum of the contributions
from three states - the 1$^{-}$ at 6.198 MeV, the 2$^{-}$ at 6.35
MeV and the 3$^{-}$ at 6.404 MeV. The red, blue and green lines represent
individual contributions of the 1$^{-}$, 3$^{-}$ and 2$^{-}$ states
respectively. The contribution of the 2$^{-}$ state was not varied
(see discussion in the text).}

\end{figure}

\begin{figure}
\includegraphics[width=8.5cm]{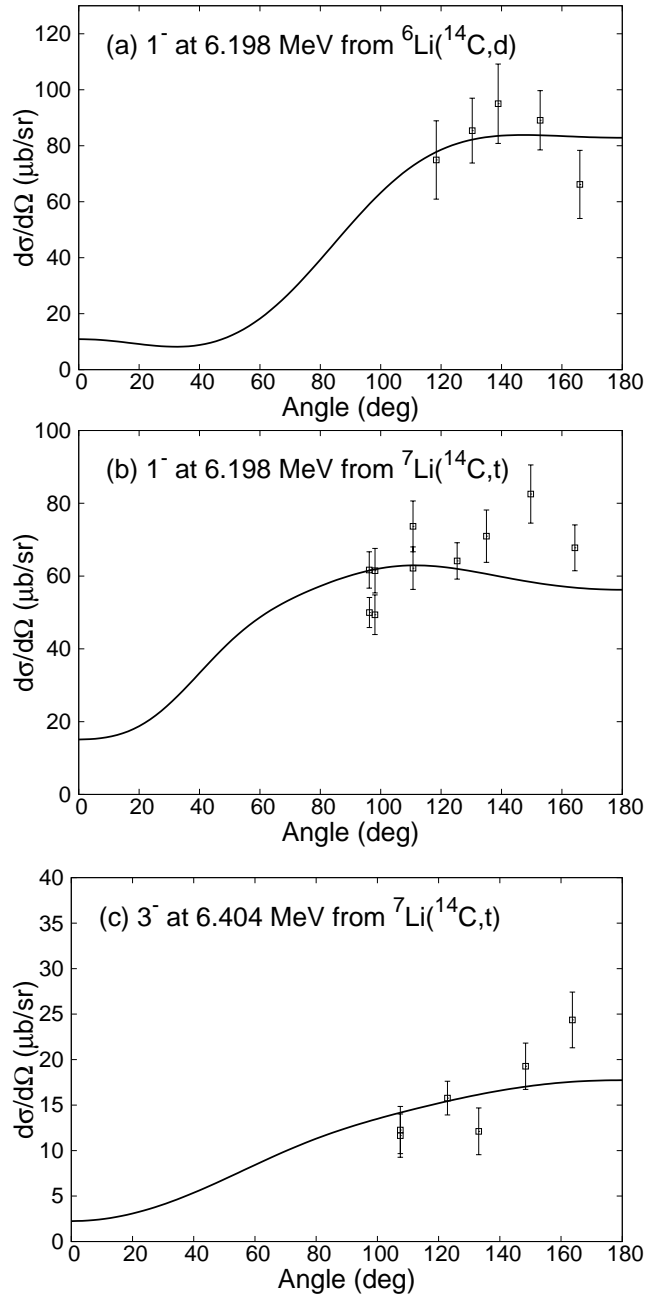}

\caption{\label{fig:Angular-distributions}Angular distributions for the 1$^{-}$
at 6.198 MeV and 3$^{-}$ at 6.404 MeV states from the $^{6}$Li($^{14}$C,d)
and $^{7}$Li($^{14}$C,t) reactions measured at 8.8 and 11.5 MeV
respectively. The solid lines correspond to the DWBA best fit.}

\end{figure}

A sample of the three states fit to the peak which corresponds to
$\sim6.3$ MeV excitation in $^{18}$O is shown in Figure \ref{fig:main-fit}.
The strengths of the 1$^{-}$ and the 3$^{-}$ states at 6.198 and
6.404 MeV are the only two free parameters. Strength of the 2$^{-}$
state was fixed as described above. The same fitting procedure was
performed for all telescopes. The angular distributions for the 6.198
MeV and 6.404 MeV states from the $^{6}$Li($^{14}$C,d) and $^{7}$Li($^{14}$C,t)
reactions are shown in Figure \ref{fig:Angular-distributions}.

\begin{table}
\caption{\label{tab:optical-model}Optical-model parameters used in analysis
of the $^{14}$C($^{7}$Li,t) and $^{14}$C($^{6}$Li,d) reactions.}

\begin{longtable}{cccccccccc}
\hline 
\toprule  & V$^{a}$  & R  & a  & W$^{a}$  & W$_{d}^{b}$  & R$'$  & a$'$  & R$_{c}$  & Ref.\tabularnewline
Channel  & (MeV)  & (fm)  & (fm)  & MeV  & MeV  & (fm)  & (fm)  & (fm)  & \tabularnewline
\hline 
$^{14}$C+$^{7}$Li  & 33.1  & 4.17  & 0.85  & 0.0  & 7.8  & 4.51  & 0.72  & 4.17  & \cite{Fortune1978}\tabularnewline
$^{18}$O+t  & 130.0  & 3.43  & 0.72  & 8.0  & 0.0  & 4.19  & 0.80  & 3.43  & \cite{Fortune1978}\tabularnewline
$^{14}$C+$^{6}$Li  & 250.0  & 3.26  & 0.65  & 0.0  & 7.5  & 3.26  & 0.65  & 4.82  & \cite{Cunsolo1981}\tabularnewline
$^{18}$O+d  & 92.92  & 2.73  & 0.81  & 0.0  & 2.5  & 3.65  & 0.71  & 3.4  & \cite{Cunsolo1981}\tabularnewline
$^{14}$C+$\alpha$  & v  & 3.53  & 0.6  &  &  &  &  & 3.53  & \cite{Buck1975}\tabularnewline
$\alpha$+d  & v  & 1.9  & 0.65  &  &  &  &  & 1.9  & \cite{Kubo1972}\tabularnewline
$\alpha$+t  & v  & 2.05  & 0.7  &  &  &  &  & 2.05  & \cite{Kubo1972}\tabularnewline
\hline
\bottomrule  &  &  &  &  &  &  &  &  & \tabularnewline
\end{longtable}

\begin{raggedright}
$^{a}$Form factor: Woods-Saxon. 
\par\end{raggedright}

\begin{raggedright}
$^{b}$Form factor: Woods-Saxon derivative, values in the table do
not include a regular factor of 4: W=4W$_{d}$. 
\par\end{raggedright}

\begin{raggedright}
v - varied to reproduce separation energies. 
\par\end{raggedright}
\end{table}

DWBA analysis of the angular distributions was performed using the
code FRESCO (version FRES 2.4) \cite{Thompson88}. The optical model
potentials parameters for the DWBA calculations were adopted from
\cite{Cunsolo1981} and \citet{Fortune1978} for the $^{6}$Li($^{14}$C,d)
and the $^{7}$Li($^{14}$C,t) reactions respectively. The shape of
the form factor potentials is from \citet{Kubo1972} and \citet{Buck1975}.
These parameters are summarised in Table \ref{tab:optical-model}.
It is important to note that both the 1$^{-}$ and the 3$^{-}$ states
were treated as bound in the DWBA analysis, even though the 3$^{-}$
is actually unbound by 177 keV. This is a good approximation since
this state decays only by $\gamma$ emission. The dependence of the
final result on the choice of the binding energy will be discussed
later in this section. The DWBA calculations were performed for beam
energies at the center of the target, 11.42 MeV for the $^{7}$Li
target measurements and 8.7 MeV for the $^{6}$Li target measurements.
The ANCs for the 1$^{-}$ and the 3$^{-}$ states were extracted from
the experimental data and the DWBA analysis.

Following the results of \cite{Mukhamedzhanov1999} the ANC is defined
in the following way. The radial overlap function of the bound state
$c$ consisting of two particles $a+b$ (c=(ab)) can be approximated
by a model wave function:\begin{equation}
I_{(ab)lj}^{c}(r)=\sqrt{S_{(ab)lj}^{c}}\psi_{nlj}^{c}(r),\label{eq:overlap}\end{equation}
 where $\psi_{nlj}^{c}(r)$ is the bound state wave function for the
relative motion of $a$ and $b$, and $S_{(ab)lj}^{c}$ is the spectroscopic
factor of the configuration $(ab)$ with the corresponding quantum
numbers $l$ and $j$ in state $c$. The asymptotic normalization
coefficient $C_{(ab)lj}^{c}$ defines the amplitude of the tail of
the radial overlap function $I_{(ab)lj}^{c}$ and at radii beyond
nuclear interaction radius $(r>R_{N})$ is given by

\begin{equation}
I_{(ab)lj}^{c}(r)\rightarrow C_{(ab)lj}^{c}\frac{W_{-\eta_{c},l_{c}+1/2}(2k_{ab}r)}{r}.\label{eq:overlap_asym}\end{equation}
 $W_{-\eta_{c},l_{c}+1/2}(2k_{ab}r)$ is the Whittaker function describing
the asymptotic behavior of the bound state wave function of two charged
particles, $k_{ab}$ is the wave number of the bound state $c$ ($k_{ab}=\sqrt{2\mu_{ab}\varepsilon_{c}}$)
and $\eta_{c}$ is the Coulomb parameter $\eta_{c}=Z_{a}Z_{b}\mu_{ab}/k_{ab}$
of the bound state $c$. The bound state wave function $\psi_{nlj}^{c}(r)$
has similar asymptotic behavior\begin{equation}
\psi_{nlj}^{c}(r)\rightarrow b_{nlj}^{c}\frac{W_{-\eta_{c},l_{c}+1/2}(2k_{ab}r)}{r},\label{eq:spwf}\end{equation}
 where $b_{nlj}^{c}$ is the single-particle ANC. From the equations
\eqref{eq:overlap}, \eqref{eq:overlap_asym} and \eqref{eq:spwf},\begin{equation}
(C_{(ab)lj}^{c})^{2}=S_{(ab)lj}^{c}(b_{nlj}^{c})^{2}\label{eq:sf}\end{equation}
 Spectroscopic factors can be extracted directly from the experimental
data \begin{equation}
\frac{d\sigma}{d\Omega}_{exp}=S_{1}S_{2}\frac{d\sigma}{d\Omega}_{DWBA},\label{eq:cs}\end{equation}
 where $S_{1}$ is the $\alpha+^{14}$C spectroscopic factor of the
corresponding state in $^{18}$O and $S_{2}$ is the $\alpha+d(t)$
spectroscopic factor of the ground state of $^{6(7)}$Li. The squares
of $\alpha+d(t)$ ANCs of the ground states of $^{6}$Li and $^{7}$Li
are well known and equal to $5.29\pm0.5$ and $12.74\pm1.1$ fm$^{-1}$
respectively \citet{blok93,Igamov2007}. Therefore, the ANCs for the
$^{18}$O states populated in the ($^{6,7}$Li,d/t) reactions can
be extracted. The R-matrix reduced width of the state can then be
determined from the ANC \cite{Mukhamedzhanov1999}\begin{equation}
\gamma_{c}^{2}=\frac{1}{2\mu_{ab}}\frac{W_{-\eta_{c},l+1/2}^{2}(2k_{ab}R)}{R}(C_{(ab)lj}^{c})^{2},\label{eq:red_width}\end{equation}
 where $R$ is the channel radius. The partial $\alpha$ width $\Gamma_{\alpha}$
of the state, which is the only missing parameter needed to determine
the astrophysically relevant resonance strength (see discussion in
section \ref{sec:introduction}), is given by equations \eqref{eq:width}
and \eqref{eq:pen} in which $\gamma_{c}^{2}$ is used instead of
$\gamma_{SP}^{2}$.

The DWBA cross section is proportional to the squares of the single
particle ANCs determined by the form factor potentials. Spectroscopic
factors are inversely proportional to the single-particle ANCs (see
eq. \eqref{eq:sf}). This makes ANCs extracted from the experimental
data using equations \eqref{eq:sf} and \eqref{eq:cs} insensitive
to the details of the form factor potentials. Since the reaction is
performed at sub-Coulomb energies the dependence of the DWBA cross
section on the parameters of the optical model potentials is minimized.

The partial $\alpha$ width of the 3$^{-}$ state at 6.404 MeV in
$^{18}$O determined from the measured cross sections using the approach
described above is $\Gamma_{\alpha}=(7.8\pm2.7)\times10^{-14}$ eV,
and a detailed analysis of the parameter dependence of the result
follows. The reduced width $\gamma_{\alpha}$ is inversely proportional
and the penetrability factor is directly proportional to the channel
radius (see Eqs. \eqref{eq:red_width} and \eqref{eq:pen}), so that
the explicit dependence of the partial $\alpha$ width on the channel
radius cancels out. Only the implicit dependence through the $kR$
argument of the Coulomb and Whittaker functions remains. The partial
$\alpha$ width of the 3$^{-}$ state calculated at the channel radius
of 4 fm is $8.2\times10^{-14}$ eV and at the channel radius of 6
fm it is $7.6\times10^{-14}$ eV. Therefore, the channel radius uncertainty
contributes to the total error budget at the level of 8$\%$. A channel
radius of 5.2 fm was adopted.

The value of the spectroscopic factor extracted from the transfer
reaction depends on the assumption of the number of radial nodes of
the bound state wave function but the value of the corresponding ANC
does not. The $\alpha$+d(t) wave function which corresponds to the
$^{6(7)}$Li ground state has one node (excluding the origin and infinity).
The number of nodes of the $\alpha$+$^{14}$C wave function for the
3$^{-}$ state in $^{18}$O is not known. From the Talmi-Moshinski
relation $2N+L=\sum(2n_{i}+l_{i})$, where the sum is over all nucleons
in the cluster, the minimum number of nodes in the $\alpha$ cluster
wave function of the 3$^{-}$ state is 2. The DWBA calculations for
2, 3 and 4 node wave functions were performed producing $\Gamma_{\alpha}=$
8.0, 7.9 and 7.8 $\times10^{-14}$ eV respectively, so a 3$\%$ uncertainty
is associated with the unknown number of nodes in the $\alpha$ cluster
wave function of the $3^{-}$ state at 6.404 MeV in $^{18}$O. Three
nodes were considered as the {}``best guess''.

The 3$^{-}$ state is unbound by 177 keV but treated as a bound state
with small binding energy. The choice of binding energy is arbitrary
and may, in principle, influence the final result. We performed calculations
for several binding energies between 200 and 10 keV. While the dependence
of the final result (value of $\Gamma_{\alpha}$) on the specific
value of the binding energy is very small there is a clear trend of
decreasing $\Gamma_{\alpha}$ with decreasing binding energy. Note,
that the penetrability factor was always calculated at resonance energy,
which is 177 keV above the decay threshold. The extrapolation of this
trend into the positive energy region suggests that the partial $\alpha$
width of a state at 177 keV in the c.m. should be reduced by about
2\% from the value obtained with 30 keV binding. The final value of
$\Gamma_{\alpha}=7.8\times10^{-14}$ eV takes this small correction
into account.

The $^{7}$Li form factor potential was adopted from \citet{Kubo1972}
(Woods-Saxon with V=-91.2 MeV, R=2.05 fm and a=0.7 fm). This potential
has been shown to reproduce t+$\alpha$ scattering phase shifts at
low energies as well as the $\alpha$-particle binding energy in $^{7}$Li
(with one node and an $\ell$=1 wave function). It was used also as
the main coupling potential for the ($^{7}$Li,t) $\alpha$ transfer
reaction. The dependence of $\Gamma_{\alpha}$ on the specific choice
of the Li form factor shape parameters was investigated. An increase
of the radius parameter from 2.05 fm to 3.7 fm (sum of rms charge
radius of $\alpha$-particle and $^{3}$H) increases $\Gamma_{\alpha}$
from 7.8 to 8.1$\times$10$^{-14}$ eV. The potential depth is always
adjusted to reproduce the $\alpha$+t binding energy with one node
and an $\ell=1$ wave function. Variation of the diffuseness parameter
of this potential within reasonable limits (from 0.5 to 0.8 fm) has
an even smaller influence on the final value of $\Gamma_{\alpha}$
($\sim2\%$).

The dependence of the final result on the specific parameters of the
$^{14}$C+$\alpha$ binding potential is also weak. The $^{18}$O
form factor potential was taken from \citet{Buck1975} (Woods-Saxon
with R=3.53 fm and a=0.6 fm). The potential depth was adjusted to
produce the desired binding energy. Variation of the radius parameter
from 3.0 fm to 4.2 fm results in $\sim4\%$ variation of the partial
$\alpha$ width and variation of the diffuseness parameter from 0.5
to 0.8 fm produces $\sim10\%$ variation. Note that the DWBA cross
section depends strongly on the specific choice of the form factor
potential parameters and varies by an order of magnitude with the
changes described above. However, the value of the ANC is not sensitive
to these variations.

The influence of the optical model potential parameters on the final
result was minimized by performing the $\alpha$-transfer reaction
at a sub-Coulomb energy. Yet, since the $^{14}$C beam energy was
close to the Coulomb barrier uncertainty in the parameters of the
optical model potentials gives largest contribution to the final uncertainty
of the $\Gamma_{\alpha}$ value of the 3$^{-}$ state. Assessment
of this uncertainty is somewhat arbitrary but was determined here
by performing a Monte Carlo variation of all the nuclear potential
parameters. This process resulted in a Gaussian distribution of DWBA
cross sections with a standard deviation of 30$\%$. The radius and
diffuseness changes to the imaginary part of the $^{14}$C+$^{7}$Li
optical potential contributed the most to the uncertainty. DWBA calculations
were performed with several different sets of $^{14}$C+$^{7}$Li
potentials taken from \cite{Poling1976,Fortune1978,Cunsolo1981}
and verified that all of these (very different) potentials give $\Gamma_{\alpha}$
well within the 30$\%$ uncertainty limit determined using the Monte
Carlo approach.

The statistical uncertainty for the $^{7}$Li($^{14}$C,t)$^{18}$O(3$^{-}$)
reaction data is 7$\%$. The combined systematic uncertainty in the
absolute cross section normalization originating from the determination
of the product of the target thickness times the solid angle is also
7$\%$. Finally, there is a 9$\%$ uncertainty in the known value
of the $\alpha+t$ ANC, 12.74$\pm1.1$ fm$^{-1}$ \cite{Igamov2007}.
Collecting all the components of the error budget yields the final
relative uncertainty of the $\Gamma_{\alpha}$ value of the 3$^{-}$
state at 6.404 MeV in $^{18}$O determined from the $^{7}$Li($^{14}$C,t)
$\alpha$-transfer reaction at sub-Coulomb energy of 35$\%$.

An ANC can only be reliably extracted from the transfer reaction data
if reaction is peripheral. Performing transfer reaction at sub-Coulomb
energy is an important factor in achieving this condition. Limiting
radius at which transfer reaction is considered in DWBA calculations
it was verified, that 95$\%$ of the transfer reaction cross section
comes from radius beyond 7 fm. This is far outside of the sum of nuclear
interaction radius of $^{7}$Li and $^{14}$C indicating that reaction
is highly peripheral.

The determination of the ANC of the 3$^{-}$ state at 6.404 MeV in
$^{18}$O was also attempted from the $^{6}$Li($^{14}$C,d) reaction
data. Unfortunately, statistics were rather low and only an upper
limit could be set on the cross section from the $^{6}$Li($^{14}$C,d)
data. Therefore, only an upper limit of $\sim10{}^{-13}$ eV can be
determined for the $\Gamma_{\alpha}$ of this state which is consistent
with the $\Gamma_{\alpha}=(7.8\pm2.7)\times10^{-14}$ eV determined
from the $^{7}$Li($^{14}$C,t) data.

The ANC for the $1^{-}$ state at 6.198 MeV and the corresponding
uncertainty was determined using the same steps as outlined above.
The only difference is that this state is bound by 29 keV with respect
to $\alpha$-decay, therefore no correction to the dependence of the
ANC on the assumed binding energy was necessary. The square of the
Coulomb modified ANC for this state was extracted from both the $^{7}$Li($^{14}$C,t)
and the $^{6}$Li($^{14}$C,d) data sets. The resulting values are
2.6$\pm$0.9 fm$^{-1}$ and 3.0$\pm$1.0 fm$^{-1}$. The two values
extracted from the different reactions agree within error. As before,
the uncertainty is dominated by the dependence on the optical model
potential parameters. The final result for the squared Coulomb modified
ANC of the 1$^{-}$ state at 6.198 MeV in $^{18}$O, determined by
combining values from both the ($^{6}$Li,d) and ($^{7}$Li,t) measurements,
is 2.8$\pm$0.7 fm$^{-1}$.

\begin{figure}
\includegraphics[width=1\columnwidth]{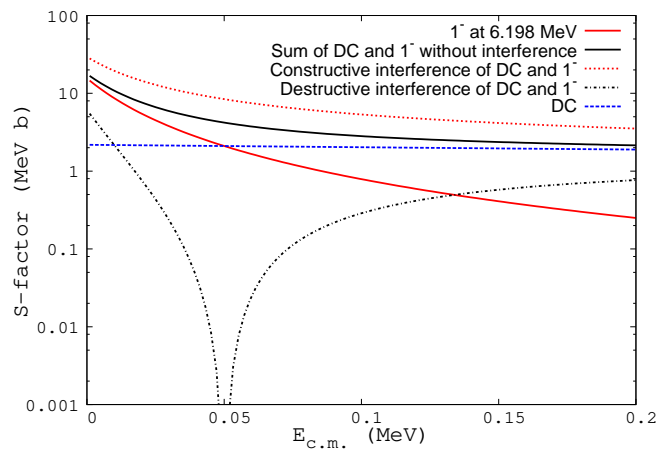}

\caption{\label{fig:S-factor}(Color online) S-factor of the $^{14}$C($\alpha$,$\gamma$)
reaction as a function of energy. The blue dashed line is the S-factor
due to direct capture (adopted from \citet{Gorres1992}), the red
solid line is the S-factor due to the sub-threshold 1$^{-}$ state
at 6.198 MeV in $^{18}$O and the black solid line is the sum of the
two without interference. The red dotted line corresponds to constructive
and the black dash-dotted line to destructive interference of the
1$^{-}$ state at 6.198 MeV with the direct capture amplitude.}

\end{figure}

\section{The $^{14}$C($\alpha$,$\gamma$) reaction rate\label{sec:rate}}

With the $\Gamma_{\alpha}$ for the 3$^{-}$ state at 6.404 MeV and
ANC for the 1$^{-}$ at 6.198 MeV in $^{18}$O, the $^{14}$C($\alpha$,$\gamma$)
reaction rate can be reliably extrapolated down to very low temperatures.
At temperatures below 1 GK four major contributors to the $^{14}$C($\alpha$,$\gamma$)
reaction rate can be identified. These are the direct capture (DC),
resonance capture through the 4$^{+}$ and 3$^{-}$ states at 7.12
and 6.404 MeV and the sub-threshold resonance capture through the
$1$$^{-}$ state at 6.198 MeV. Direct $^{14}$C($\alpha$,$\gamma$)
capture was measured by \citet{Gorres1992} in the energy range from
1.14 to 2.33 MeV and was extrapolated to lower energies using the
following parametrization of the astrophysical S-factor

\begin{equation}
S(E)=2.18-1.60E+0.82E^{2}.\end{equation}

The resonance strength of the 4$^{+}$ state was also determined by
direct measurements in \cite{Gorres1992} ($\omega\gamma=0.46\pm0.08$
eV). The resonance strength of the 3$^{-}$ state determined through
the ANC in this work is $\omega\gamma=7\times\Gamma_{\alpha}=(5.5\pm1.9)\times10^{-13}$
eV. The resonance reaction rate was calculated as in \citet{Fowler1975}

\begin{equation}
\begin{array}{ll}
N_{A}<\sigma\nu>_{R}= & 1.54\times10^{11}(\mu T_{9})^{-3/2}\times\omega\gamma\times\\
 & \times\exp\left(\frac{-11.605E_{r}}{T_{9}}\right),\end{array}\end{equation}

\noindent where $\mu$ is a reduced mass and T$_{9}$ is the temperature
in GK.

\begin{figure}
\includegraphics[width=1\columnwidth]{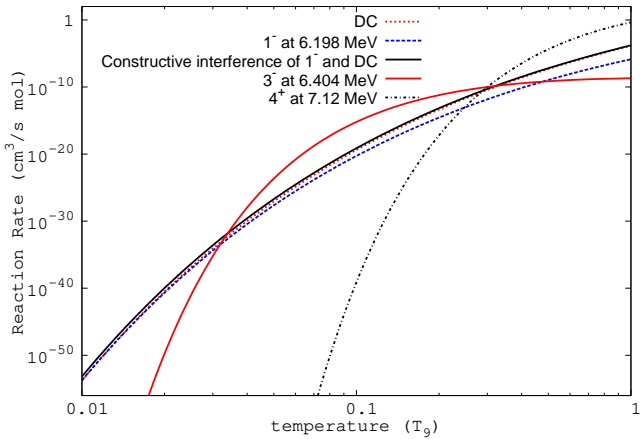}

\caption{\label{fig:rates}(Color online) The $^{14}$C($\alpha$,$\gamma$)
reaction rate due to resonant and non-resonant capture. Resonant capture
due to the 4$^{+}$ state at 7.12 MeV and 3$^{-}$ state at 6.404
MeV are shown as the black dash-dotted and solid red curves, respectively.
Direct capture is the red dotted curve, capture due to the 1$^{-}$
sub-threshold resonance at 6.198 MeV is the blue dashed curve.}

\end{figure}

The astrophysical S-factor for the 1$^{-}$ sub-threshold resonance
was calculated from the Coulomb modified ANC using the approach outlined
in \cite{Mukhamedzhanov1999} as

\begin{equation}
S(E)\approx(2L+1)\pi^{2}\frac{k_{c}}{\mu^{2}}\frac{\eta_{c}^{2L+1}}{(E+\varepsilon_{c})^{2}}\Gamma_{\gamma}\times|\widetilde{C}_{1^{-}}|^{2},\end{equation}

\noindent where $|\widetilde{C}_{1^{-}}|^{2}$ is the square of the
Coulomb modified ANC (ANC divided by the corresponding Gamma function,
$\Gamma(L+1+\eta_{c})$) of the 1$^{-}$ state, $k_{c}$ is the wave
number of the 1$^{-}$ state $k_{c}=\sqrt{2\mu\varepsilon_{c}}$,
$\varepsilon_{c}$ is the binding energy, $\Gamma_{\gamma}$ is the
partial $\gamma$ width of the 1$^{-}$ state calculated from the
known mean lifetime, $\Gamma_{\gamma}=(1.8\pm0.3)\times10^{-7}$ MeV.
The combined relative uncertainty of the $S(E)$ value is 30$\%$.
The S-factor due to direct capture is compared to the S-factor due
to the 1$^{-}$ sub-threshold resonance in Figure \ref{fig:S-factor}.
It is important to note that at energies below 50 keV the contribution
to the S-factor from the 1$^{-}$ sub-threshold resonance is larger
than from the direct capture. Based on the results of \cite{Gorres1992}
direct capture is mainly due to the p-wave. Therefore, the direct
capture amplitude may interfere with the 1$^{-}$ sub-threshold resonance
amplitude. The S-factors due to constructive and destructive interference
of these amplitudes are shown in Fig. \ref{fig:S-factor} as red dotted
and black dash-dotted curves, respectively. Obviously, interference
significantly amplifies the importance of the 1$^{-}$ state. Unfortunately,
the sign of this interference is not known and both cases will have
to be considered. The reaction rates due to direct capture and capture
to the sub-threshold 1$^{-}$ state at 6.198 MeV in $^{18}$O were
calculated by numerical integration of:

\begin{equation}
\begin{array}{ll}
<\sigma\nu>_{NR}= & \sqrt{\frac{8}{\mu\pi}}\left(kT\right)^{-3/2}\int S(E)\times\\
 & \times\exp(-2\pi\eta)\exp\left(-\frac{E}{kT}\right)dE.\end{array}\end{equation}

The reaction rates due to resonance and non-resonance capture are
shown in Figure \ref{fig:rates} as a function of temperature. Three
temperature regions can be identified. At T>0.3 GK the $^{14}$C($\alpha$,$\gamma$)
reaction rate is dominated by the $4^{+}$ state at 7.12 MeV. The
4$^{+}$ resonance strength was measured directly \cite{Gorres1992,Gai1987}
and the uncertainty of the $^{14}$C($\alpha$,$\gamma$) reaction
rate in this region is $17\%$. In the temperature range between 0.03
and 0.3 GK the 3$^{-}$ state at 6.404 MeV dominates. This temperature
range is of particular interest for helium accreting white dwarfs
and AGB stars. The 3$^{-}$ resonance strength was determined in this
work with an uncertainty of $35\%$, which determines the new uncertainty
for the $^{14}$C($\alpha$,$\gamma$) reaction rate in the temperature
range of 0.03-0.3 GK. Below 0.03 GK, direct capture and capture due
to the sub-threshold 1$^{-}$ state at 6.198 MeV dominate. While direct
capture was extrapolated from direct measurements in \citet{Gorres1992}
and the S-factor due to the 1$^{-}$ sub-threshold resonance was determined
in this work with an uncertainty of 30$\%$ the reaction rate at this
low energy region is still uncertain by two orders of magnitude due
to the unknown interference between the direct and the 1$^{-}$ sub-threshold
resonance capture amplitudes (see Figure \ref{fig:S-factor}).

\begin{figure}
\includegraphics[width=1\columnwidth]{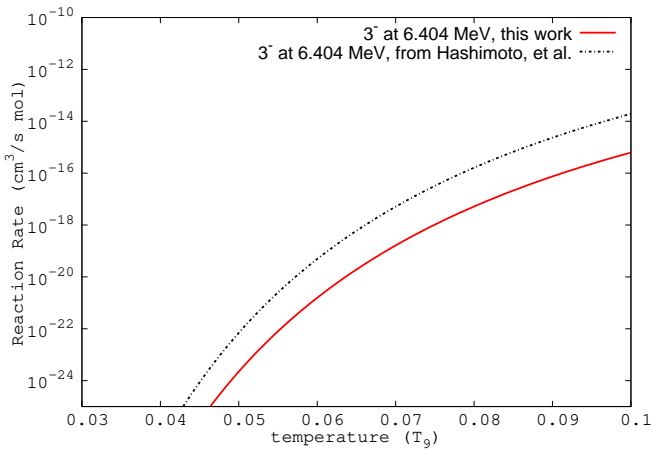}
\caption{\label{fig:Hashimoto} (Color online) Comparison of the new $^{14}$C($\alpha$,$\gamma$)
reaction rate (red solid curve) in the temperature range most relevant
for the helium accreting white dwarfs to the rate used by Hashimoto,
et al., \cite{Hashimoto1986} (black dash-dotted curve).}
\end{figure}

The new $^{14}$C($\alpha$,$\gamma$) reaction rate may have the
most profound effect on the evolution of accreting helium white dwarfs.
It was suggested by \citet{Hashimoto1986} that under certain conditions
the $^{14}$N(e$^{-}$,$\nu$)$^{14}$C($\alpha$,$\gamma$)$^{18}$O
(NCO) reaction dominates over the 3$\alpha$ reaction and triggers
a helium flash at earlier times and lower density values than the
3$\alpha$ process would otherwise. However, this suggestion is sensitive
to the actual $^{14}$C($\alpha$,$\gamma$) reaction rate. Various
accretion rates have been considered in \citet{Hashimoto1986} and
it was concluded that the effect of the NCO reaction is larger for
faster accretion. Following the evolutionary path of the central density
and temperature of the helium accreting white dwarf with the accretion
rate of 10$^{-8}$ M$_{\bigodot}$ yr$^{-1}$ it was found (see Fig.
5 in \citet{Hashimoto1986}) that $^{14}$C burning is ignited when
the central temperature reaches 0.066 GK while the 3$\alpha$ reaction
does not begin until 0.080 GK. This prediction was based on the hypothetical
$^{14}$C($\alpha$,$\gamma$) reaction rate which turned out to be
a factor of 30 higher than the rate determined from our measurements.
(Note that at these temperatures the reaction rate is dominated by
the resonance capture due to the 3$^{-}$ state at 6.404 MeV). Comparison
of the reaction rate in the temperature range of interest for helium
accreting white dwarfs used by Hashimoto and the one determined from
our experimental data is shown in Figure \ref{fig:Hashimoto}. The
new, much lower reaction rate would change the NCO ignition temperature
to higher values. An estimate of the new ignition temperature based
on equality of the corresponding $^{14}$C($\alpha$,$\gamma$) reaction
rates gives a new value of about 0.075 GK which is very close, but
still lower than the ignition temperature for the 3$\alpha$ reaction.
Obviously, detailed calculations with the new $^{14}$C($\alpha$,$\gamma$)
reaction rate are necessary to know whether the NCO reaction has an
effect on the evolution of the helium accreting white dwarf. However,
it is clear that the new, significantly reduced reaction rate casts
serious doubt on the effectiveness of the NCO reaction chain as a
trigger for helium flashes compared to the 3$\alpha$ reaction.

It was found in \citet{Lugaro2004} that uncertainties in the $^{14}$C($\alpha$,$\gamma$)
and the $^{19}$F($\alpha$,p) reaction rates are the main contributing
factors to the uncertainty of the production of $^{19}$F in AGB stars.
The authors of \citet{Lugaro2004} cite 5 orders of magnitude uncertainty
in the $^{14}$C($\alpha$,$\gamma$) reaction rate due to the unknown
spectroscopic factor of the $3^{-}$ state at 6.404 MeV. This uncertainty
is eliminated now. Surprisingly, the {}``recommended'' $^{14}$C($\alpha$,$\gamma$)
reaction rate used in \citet{Lugaro2004}, which is based on the assumption
that the spectroscopic factor of the 3$^{-}$ state is 0.01 is almost
exactly (within experimental uncertainty of this work) the same as
the reaction rate determined from our measurements. This fortuitous
coincidence means that the yields of $^{19}$F calculated in \citet{Lugaro2004}
with the {}``recommended'' $^{14}$C($\alpha$,$\gamma$) reaction
rate are accurate and no new calculations are necessary. Of course,
the uncertainty due to the $^{19}$F($\alpha$,p) reaction rate still
remains and will have to be addressed.

Finally, with the uncertainty for the $^{14}$C($\alpha$,$\gamma$)
reaction rate reduced to just 35$\%$ in the temperature range of
interest for the core helium flashes in low-mass stars the debate
over possible importance of the NCO reaction chain as a trigger can
hopefully be settled.

\section{Conclusion\label{sec:Conclusions}}

The $^{14}$C($\alpha$,$\gamma$) reaction rate was studied using
an indirect approach. The asymptotic normalization coefficients (ANCs)
for the 1$^{-}$ state at 6.198 MeV and 3$^{-}$ at 6.404 MeV in $^{18}$O
were measured in the $\alpha$-transfer reactions ($^{7}$Li,t) and
($^{6}$Li,d) on $^{14}$C. The measurements were performed at sub-Coulomb
energies to minimize the dependence of the final result on the optical
model potential parameters. The extraction of ANCs instead of spectroscopic
factors significantly reduced the dependence on the form-factor potential
parameters and the number of nodes assumed in the $\alpha$-$^{14}$C
wave function. The resonance strength of the 3$^{-}$ state was determined
from the ANC to be $\omega\gamma=(5.5\pm1.9)\times10^{-13}$ eV. It
was found that the resonance capture due to the 3$^{-}$ state dominates
the reaction rate in the temperature range of 0.03<T$_{9}$<0.3 which
is the most relevant temperature range for the $^{19}$F nucleosynthesis
in AGB stars, evolution of helium accreting white dwarfs and core
helium flashes of low-mass stars. At temperatures below 0.03 GK the
$^{14}$C($\alpha$,$\gamma$) reaction rate is determined by the
interplay between the direct capture and capture due to the sub-threshold
1$^{-}$ resonance at 6.198 MeV. The S-factor due to the 1$^{-}$
sub-threshold state was determined in this work from the ANC and the
direct capture S-factor was suggested in \cite{Gorres1992}. In spite
of this knowledge, the reaction rate at the temperatures below 0.03
GK is still uncertain by two orders of magnitude due to the unknown
interference (constructive or destructive) between the two amplitudes.

The new $^{14}$C($\alpha$,$\gamma$) reaction rate is a factor of
30 lower than the one used in \cite{Hashimoto1986}. Therefore, the
importance of the NCO chain as a trigger for helium flashes in helium
accreting white dwarfs suggested in \cite{Hashimoto1986} is reduced,
if not eliminated all together. The {}``recommended'' $^{14}$C($\alpha$,$\gamma$)
reaction rate used in \cite{Lugaro2004} for $^{19}$F nucleosynthesis
calculation in AGB stars fortuitously coincides with that determined
from our experimental data and the uncertainty of the $^{19}$F production
in AGB stars associated with the $^{14}$C($\alpha$,$\gamma$) reaction
rate is now eliminated. We hope that the new information on the $^{14}$C($\alpha$,$\gamma$)
reaction rate will help to resolve the question of whether or not
the NCO reaction chain can trigger helium flashes in cores of low-mass
stars before the 3$\alpha$ reaction does.

Authors are grateful to Peter Hoeflich and Akram Mukhamedzhanov for
enlightening discussions and acknowledge support of National Science
Foundation under grant number PHY-456463.

\bibliographystyle{apsrev} \bibliographystyle{apsrev}
\bibliography{references}

\end{document}